\documentclass[preprint,showpacs,preprintnumbers,amsmath,amssymb,aps]{revtex4}

\usepackage{graphicx}
\usepackage{dcolumn}
\usepackage{bm}

\begin{document}

\preprint{Phys. Rev. C}

\title{Ratio of $\sigma_L$/$\sigma_T$ for $p(e,e'K^+)\Lambda$ Extracted
from Polarization Transfer}

\author{Brian A. Raue}
 \email{baraue@fiu.edu}
\affiliation{Florida International University, Miami, FL 33199}

\author{Daniel S. Carman}
 \email{carman@ohio.edu}
\affiliation{Ohio University, Athens, OH 45701}

\date{\today}

\begin{abstract}
The ratio of longitudinal to transverse structure functions,
$\sigma_L/\sigma_T$, has been extracted from recent beam-recoil
transferred polarization data for the $p(\vec e,e'K^+)\vec\Lambda$
reaction.  Results have been obtained for $W$=1.72, 1.84, and 1.98~GeV
at an average $Q^2$ of 0.77, 0.69, and 0.61~GeV$^2$, respectively.
Our results indicate a ratio that is systematically slightly smaller than
previously published results using a Rosenbluth separation.
\end{abstract}

\maketitle

\section{Introduction}
\label{sec:intro}

Data taken in Hall B at Jefferson Laboratory have recently been published 
on polarization transfer in the reaction $p(\vec{e},e'K^+)\vec{\Lambda}$
\cite{carman03}.  While these data have been used to shed light on the 
$s\bar{s}$ quark pair creation operator in the associated strangeness
production reaction, they cannot yet provide direct constraints on the
isobar models commonly employed to describe the reaction mechanism
\cite{pcomm}.  These phenomenological models~(e.g.\cite{BM00,J02}) rely on 
fitting the available data to provide constraints on the contributing
intermediate-state resonant and non-resonant processes in the $s$, $t$, and
$u$ reaction channels.  The models differ in the set of specific resonant
states included, as well as in their treatment of hadronic form factors and
the restoration of gauge invariance.  However, due to the sparsity of data
for this reaction and the large number of parameters in the models, the
contributions to the intermediate state remain largely unconstrained and,
therefore, are highly uncertain.

The new polarization transfer data are difficult to include in fitting the
isobar model parameters. This is due to the fact that, by necessity, these
data were averaged over a large range in momentum transfer $Q^2$ (0.3 to
1.5~GeV$^2$) and a large range in the invariant energy $W$ ($\sim$200~MeV
bins).  The data have also been averaged over all $\Phi$, the angle between
the electron scattering plane and the $K^+\Lambda$ hadronic reaction plane
(Fig.~\ref{fig-kin}).  One could expect significant variations in the model
parameters over such large kinematic ranges.  This means that including
these data, say at their central kinematic values, into a global refit of
the parameters is improper.  Therefore these data have been employed only
as a cross check of the model parameters based on fits to other
measured observables.

The transferred polarization data can, however, provide useful new direct
constraints to the models when they are used to extract the ratio of
longitudinal to transverse structure functions, $R_{\sigma} =
\sigma_L/\sigma_T$.  This ratio has been previously measured using a
Rosenbluth separation \cite{Bebek,Niculescu,Mohring03}.  The results
presented here provide a means of extracting $R_{\sigma}$ that is less
prone to systematic uncertainties than the Rosenbluth method and give
results that are systematically slightly smaller.  

\section{Formalism}
\label{sec:formalism}

Following the notation of Ref.~\cite{Knochlein}, the most general form 
for the virtual photoabsorption cross section in the center-of-mass 
frame (c.m.) from an unpolarized target, allowing for both a polarized
electron beam and recoil hyperon, is given by:
\begin{eqnarray}
\label{csec1}
\frac{d \sigma_v}{d\Omega_K^*}\!\!\! &=& \!\!\! {\cal{K}} 
   \!\!\!\!\!\!\!\sum_{\beta=0,x',y',z'} \!\!\!\!\!
   \Bigl( R_T^{\beta 0} + \epsilon_L R_L^{\beta 0} + c_+(^c\!R_{LT}^{\beta 0} 
   \cos \Phi + \!\!^s\!R_{LT}^{\beta 0} \sin \Phi) \nonumber \\ 
 &+& \epsilon (^c\!R_{TT}^{\beta 0} 
  \cos 2 \Phi + ^s\!\!R_{TT}^{\beta 0} \sin 2 \Phi) \nonumber \\
+ h \!\!\!&c_-& \!\!\!(^c\!R_{LT'}^{\beta 0} \cos \Phi 
   + ^s\!\!R_{LT'}^{\beta 0}\sin \Phi)
+ h c_0 R_{TT'}^{\beta 0} \Bigr).
\end{eqnarray}

\begin{figure}[htbp]
\vspace{4.5cm}
\includegraphics{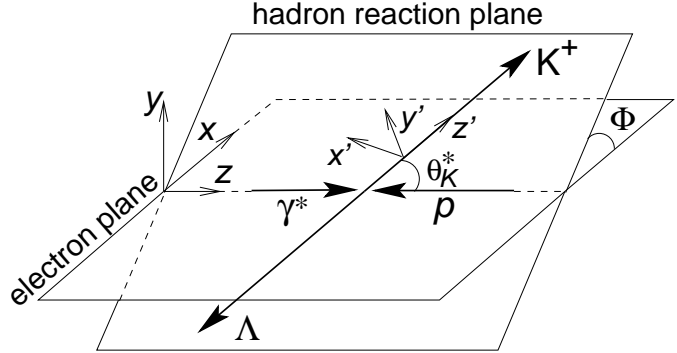} 
 \caption[]{Kinematics for $K^+\Lambda$ electroproduction showing
  angles and polarization axes in the center-of-mass reference frame.}
 \label{fig-kin}
\end{figure}

\noindent
The $R_i^{\beta \alpha}$ are the transverse, longitudinal, and
interference response functions that relate to the underlying hadronic
current and implicitly contain the $\Lambda$ polarization.  The sum
over $\beta$ includes contributions from the hyperon polarization with
respect to the $(x',y',z')$ axes.  In this system, $\hat{z}'$ is along
the outgoing $K^+$ direction, $\hat{y}'$ is normal to the hadronic
reaction plane, and $\hat{x}'= \hat{y}' \times \hat{z}'$
(Fig.~\ref{fig-kin}).  The $\beta=0$ terms account for the unpolarized
response and $\alpha$=0 implies an unpolarized target.  The response
functions denoted by $R_{LT'}$ and $R_{TT'}$ depend on the electron
beam helicity $h$.  The left superscripts on the response functions,
$c$ or $s$, indicate whether the term multiplies a sine or cosine
term, respectively.

The kinematic terms are defined by $c_{\pm} = \sqrt{2 \epsilon_L (1
\pm \epsilon)}$ and $c_0 = \sqrt{1- \epsilon^2}$, with the transverse
and longitudinal polarization of the virtual photon defined
respectively as
$\epsilon=\left[1+2(1+{\nu^2/Q^2})\tan^2{\theta_e/2}\right]^{-1}$ and
$\epsilon_L=\epsilon Q^2/(k_{\gamma}^{c.m.})^2$.  Here $\theta_e$ is
the electron scattering angle, $Q^2$ is the negative of the
four-momentum-transfer squared, $\nu$ is the virtual photon energy,
and $k_{\gamma}^{c.m.}$ is the virtual-photon c.m. momentum.  The
leading factor ${\cal{K}}= \vert \vec{q}_K\vert / k_{\gamma}^{c.m.}$,
where $\vec{q}_K$ is the kaon c.m. momentum.

Using Eq.(\ref{csec1}), the helicity-dependent hyperon polarization
components, $P'$, in the $(x',y',z')$ system are given
by~\cite{carman03,schmieden00}:
\begin{eqnarray}
\label{ptran1}
\sigma_0 P_{x'}' &=& {\cal{K}} ( c_-\,^c\!R_{LT'}^{x'0} \cos{\Phi} 
      + c_0 R_{TT'}^{x'0}), \nonumber \\ 
\sigma_0 P_{y'}' &=& {\cal{K}} c_-\,^s\!R_{LT'}^{y'0} \sin{\Phi},  \\
\sigma_0 P_{z'}' &=& {\cal{K}} ( c_-\,^c\!R_{LT'}^{z'0} \cos{\Phi} 
      + c_0 R_{TT'}^{z'0}).\nonumber
\end{eqnarray}

\noindent
Here, $\sigma_0$ represents the unpolarized part of the cross section,
which can be defined in terms of either response functions $R_i$ or
structure functions $\sigma_i$ as:
\begin{eqnarray}
\label{csec3}
\sigma_0 \equiv \frac{d\sigma_v}{d\Omega_K^*} &=& {\cal{K}}
   \left[ R_T^{00} + \epsilon_L R_L^{00} 
 + \sqrt{2\epsilon_L(1+\epsilon)}\, ^c\!R_{LT}^{00} \cos{\Phi}
 + \epsilon ^c\!R_{TT}^{00}\cos{2\Phi} \right] \nonumber \\
 &=& \sigma_T + \epsilon \sigma_L + \sqrt{2\epsilon(1+\epsilon)}\sigma_{LT}\cos{\Phi}
   + \epsilon \sigma_{TT} \cos{2\Phi},
\end{eqnarray}
where we note that $\sigma_T ={\cal{K}}R_T^{00}$ and
$\epsilon\sigma_L = {\cal{K}}\epsilon_L R_L^{00}$. 


The transferred polarization can also be defined in the $(x,y,z)$
coordinate system where $\hat{z}$ is along the virtual photon direction,
$\hat{y}$ is normal to the electron scattering plane, and $\hat{x}=\hat{y}
\times \hat{z}$ (see Fig.~\ref{fig-kin}).  The components of $P'$ in the 
$(x,y,z)$ system are related to those in the $(x',y',z')$ system by a 
simple rotation and are given by:
\begin{eqnarray}
\label{ptran2}
P'_x &=&  P'_{x'} \cos{\Phi} \cos{\theta_K^*} - P'_{y'} \sin{\Phi} 
+ P'_{z'} \cos{\Phi} \sin{\theta_K^*} \nonumber \\
P'_y &=&  P'_{x'} \sin{\Phi} \cos{\theta_K^*} + P'_{y'} \cos{\Phi}
+P'_{z'} \sin{\Phi} \sin{\theta_K^*} \\
P'_z &=& -P'_{x'} \sin{\theta_K^*} + P'_{z'} \cos{\theta_K^*}, \nonumber 
\end{eqnarray}
where $\theta_K^*$ is the $K^+$ c.m. polar angle defined in
Fig.~\ref{fig-kin}.

If the transferred polarization components are now integrated over 
all $\Phi$ (calling these components ${\cal P}'$), Eq.(\ref{ptran1})
simplifies to:
\begin{eqnarray}
\label{eqn_Pxzp}
{\cal P}_{x'}' = {\frac{c_0 R_{TT'}^{x'0}}{R_T^{00}+\epsilon_LR_L^{00}}},
\hskip 1.0cm {\cal P}_{z'}' = {\frac{c_0 R_{TT'}^{z'0}}{R_T^{00}
+\epsilon_LR_L^{00}}},
\end{eqnarray}
and Eq.(\ref{ptran2}) simplifies to:
\begin{eqnarray}
\label{eqn_Pxz}
{\cal P}_x' &=& {\frac{c_- (^c\!R_{LT'}^{x'0} \cos{\theta_K^*} - R_{LT'}^{y'0} 
         + ^s\!\!R_{LT'}^{z'0} \sin{\theta_K^*})}{R_T^{00}+\epsilon_LR_L^{00}}},
  \nonumber \\
{\cal P}_z' &=& {\frac{c_0 (-R_{TT'}^{x'0} \sin{\theta_K^*} + R_{TT'}^{z' 0} 
   \cos{\theta_K^*})}{R_T^{00}+\epsilon_LR_L^{00}}}.
\end{eqnarray}
The components of ${\cal P}'$ along the $\hat y$ and $\hat y'$ axes
are identically zero from the integration of Eqs.(\ref{ptran1}) and
(\ref{ptran2}) over $0\leq\Phi\leq 2\pi$. Such an integration was
performed on the polarization transfer data of Ref.~\cite{carman03} in
which acceptance corrections were first applied to raw yields before
summing over all $\Phi$ angles.  This had the effect of improving
statistical uncertainties on the measured transferred polarizations.
It also provided a systematics check in that the $\hat y$ and $\hat
y'$ components of the polarization were found to be zero to within
statistical uncertainties.

Concentrating now on the $z'$ and $z$ components in parallel or
anti-parallel kinematics ($\cos\theta_K^*=\pm 1$),
Eqs.(\ref{eqn_Pxzp}) and (\ref{eqn_Pxz}) reduce to:
\begin{equation} 
\label{eqn_Pz1}
 {\cal P}'_{z'} = \pm{\cal P}'_z = \pm {\frac{c_0R_{TT'}^{z' 0}} {R_T^{00} 
 + \epsilon_L R_L^{00}}} = \pm \frac{c_0 R_{TT'}^{z' 0}}{\sigma_u/{\cal{K}}},
\end{equation}
where the plus (minus) sign is associated with the parallel (anti-parallel) 
kinematics case and $\sigma_u = \sigma_T + \epsilon \sigma_L$.

The response functions of Eqs.(\ref{eqn_Pxzp}) and (\ref{eqn_Pxz}) can be 
written in terms of the CGLN amplitudes~\cite{chew} as shown in 
Ref.~\cite{Knochlein} as:
\begin{eqnarray}
\label{eqn-Rttx}
 \hskip -0.25in
 R_{TT'}^{x' 0} &=& \sin\theta_K^* {\rm Re}\left[-|F_1|^2 + |F_2|^2 +
F_2^*F_3 - F_1^*F_4 + \cos\theta_K^*(F_2^*F_4 - F_1^*F_3)\right],\\
\label{eqn-Rttz}
 \hskip -0.25in
 R_{TT'}^{z' 0} &=& {\rm Re}\left[-2F_1^*F_2 +
 \cos\theta_K^*(|F_1|^2+|F_2|^2) -
 \sin^2\theta_K^*(F_1^*F_3+F_2^*F_4)\right],\\
\label{eqn-Rt}
 R_T^{00} &=&|F_1|^2+|F_2|^2 + {\frac{\sin^2\theta_K^*} 2}(|F_3|^2+|F_4|^2)
 \nonumber \\
 & &+{\rm Re}\left[\sin^2\theta_K^*(F_2^*F_3+F_1^*F_4
 +\cos\theta_K^*F_3^*F_4) - 2\cos\theta_K^*F_1^*F_2\right],\\
\label{eqn-Rl}
 R_L^{00} &=&{\rm Re}\left[|F_5|^2+|F_6|^2+2\cos\theta_K^*F_5^*F_6\right].
\end{eqnarray}
For the case of $\theta_K^*=0$, these forms simplify to: 
\begin{equation}
\label{eqn-Rtt2}
 \hskip -0.25in
 R_{TT'}^{x' 0} = 0,~R_{TT'}^{z' 0}=R_T^{00}=|F_1-F_2|^2,  \ \ 
{\rm and}\ \ R_L^{00}=|F_5+F_6|^2.
\end{equation}
Combining the result of Eq.(\ref{eqn-Rtt2}) with Eq.(\ref{eqn_Pz1}) we obtain
for $\theta_K^*=0$:
\begin{equation}
 \label{eqn-Pz}
 {\cal P}'_{z'}={\cal P}_z'= {\frac{c_0R_T^{00}} 
  {R_T^{00}+\epsilon_L R_L^{00}}}.
\end{equation}
This expression can then be inverted to determine
the ratio of the longitudinal to transverse response
functions~\cite{schmieden00}, or alternatively, the ratio of
$R_{\sigma}=\sigma_L$/$\sigma_T$ as:

\begin{equation}
 \label{eqn-ratio2}
 R_\sigma={\frac{\sigma_L}{\sigma_T}}=
  {\frac{1}{\epsilon}}\left({\frac{c_0}{{\cal P}'_{z'}}}-1\right).
\end{equation}

\section{Extraction of $R_\sigma$}
\label{sec:extraction} 

\begin{figure}[thb]
\vspace{9.0cm}
\includegraphics{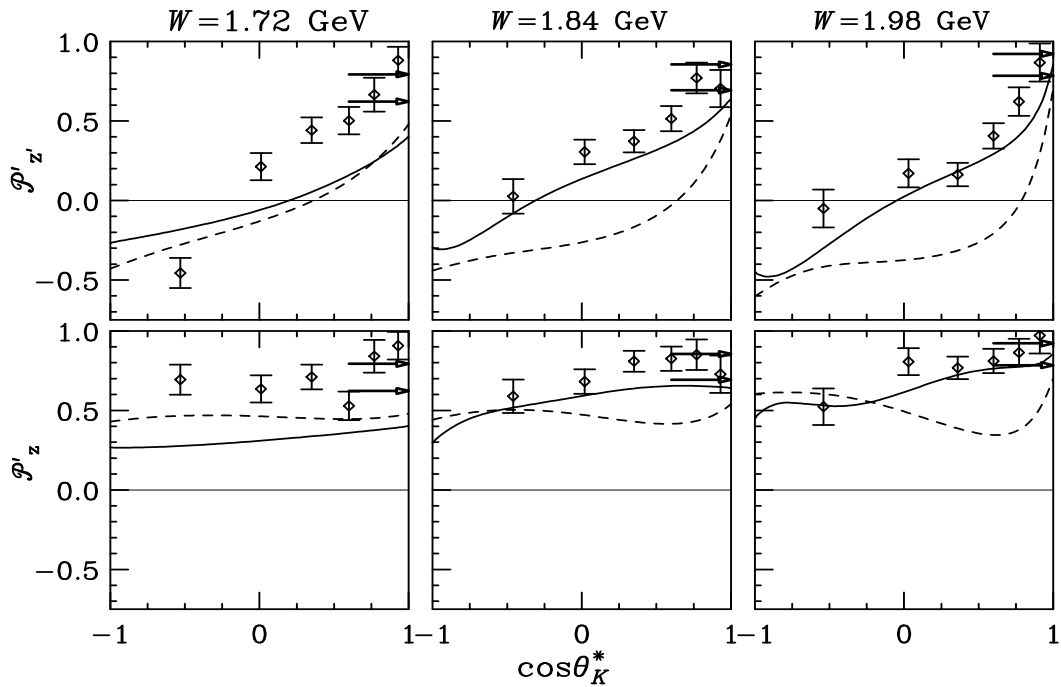} 
 \caption[]{Transferred $\Lambda$ polarization projected along the
 $z'$ (top) and $z$ (bottom) axes versus $\cos\theta^*_K$ from
 Ref.~\cite{carman03}. The data were summed over $Q^2$ and $\Phi$ and
 are shown for three average values of $W$ as indicated.  The solid
 and dashed curves correspond to the hadrodynamic models of
 Refs.~\cite{BM00} and \cite{J02}, respectively.  The upper arrow in
 each plot indicates the maximum physically allowable value of the
 polarization at $\cos\theta_K^*=1$, while the lower arrows indicate
 the polarization corresponding to $R_\sigma=0.45$.} 
\label{fig-Pz1}
\end{figure}

Figure~\ref{fig-Pz1} reproduces the results from Ref.~\cite{carman03}
along with sample model calculations. The kinematic values and the
most forward-angle data points are given in Table~\ref{tab-rawPz}. The
values of $\langle W\rangle$ and $\langle Q^2\rangle$ given in the
table and the values of $\cos\theta_K^*$ shown in the figure are the
average values over the kinematic bin and were determined by the
distribution of the acceptance-corrected data within the bin.  The 
curves correspond to the hadrodynamic models of Refs.~\cite{BM00} (solid) 
and \cite{J02} (dashed) and have been averaged over the kinematical bins. 
It is clear that the models badly miss in predicting polarization at 
nearly all values of $\cos\theta_K^*$ and, therefore, must fail to predict
$R_\sigma$.

Since both $R_L^{00}$ and $R_T^{00}$ must be positive definite, the ratio
$R_\sigma\ge 0$. Thus, Eq.(\ref{eqn-ratio2}) gives a maximum value of ${\cal
P}'_{z'}={\cal P}'_{z}=c_0$ at $\cos\theta_K^*=1$.  This limiting value is
shown by the upper arrow in each plot of Fig.~\ref{fig-Pz1}.  This arises
from the fact that these response functions can be written in terms of
absolute squares of helicity amplitudes, as shown in Ref.~\cite{Knochlein},
and must therefore always be greater than zero.

\begin{table}[htb]
 \begin{center}
 \begin{tabular}{|c|c|c|c|c|c|}\hline
 $\langle W\rangle$ GeV & $\langle Q^2\rangle$ GeV$^2$  
  & $\langle \epsilon\rangle$ & ${\cal P}'_{z'}\pm\Delta{\cal P}'_{z'}$  
  & ${\cal P}'_z\pm\Delta{\cal P}'_z$ & ${\cal P}'_{z'}(x=1)$
  \\ \hline\hline
 1.72 & 0.77 & 0.619 & 0.881$\pm$0.086 & 0.907$\pm$0.087 & 0.62 \\ \hline
 1.84 & 0.69 & 0.518 & 0.704$\pm$0.116 & 0.728$\pm$0.118 & 0.69 \\ \hline
 1.98 & 0.61 & 0.388 & 0.868$\pm$0.120 & 0.971$\pm$0.114 & 0.78 \\ \hline
 \end{tabular}
 \end{center}
 \caption{\label{tab-rawPz} Transferred polarization data at
 $\cos\theta_K^*\! =\! 0.93$ from Ref.~\cite{carman03} at the average kinematic
 quantities shown. The $Q^2$ range extends from 0.35 up to 1.4, 1.2, and
 1.0 GeV$^2$ for the kinematic bins, respectively. The $W$ ranges are 1.60
 to 1.78 GeV, 1.78 to 1.90 GeV, and 1.90 to 2.15 GeV. The last column gives
 the values of ${\cal P}'_{z'}$ and ${\cal P}'_z$ at
 $x\equiv\cos\theta_K^*\! =\! 1$ for a ratio of $R_\sigma=0.45$--the
 average value from Ref.~\cite{Mohring03}.}
\end{table}

The small-angle data of Ref.~\cite{carman03} cover an angular bin of $0.8
\leq \cos\theta_K^*\leq 1$ and therefore do not give direct access to
${\cal P}'_{z'}$ or ${\cal P}'_z$ at $\cos\theta_K^*=1$.  However, the
values of the polarization in the smallest angle bins and the trends of the
data suggest some small difference from what is expected given the recent
$R_\sigma$ results of Mohring {\it et al.}~\cite{Mohring03}. The three
Mohring data points at $Q^2$=0.52, 0.75, and 1.00~GeV$^2$ and $W$=1.84~GeV
give an approximate average value of $R_\sigma=0.45$.  This would give
polarizations at $\cos\theta_K^*=1$ shown by the lower arrows in the plots
of Fig.~\ref{fig-Pz1} and also given in the last column of
Table~\ref{tab-rawPz}.  The Mohring results suggest a somewhat lower value
for the polarization at $\cos\theta_K^*=1$ than the trends of the
polarization data at both $W=1.72$ and 1.98~GeV indicate.  One should
probably not be too surprised by the apparent discrepancy between the 
polarization results at $W=1.72$ GeV and the polarization value suggested
by the Mohring data. This bin covers the threshold region where the
$S_{11}$(1650), $P_{11}$(1710), and $P_{13}$(1720) resonances are expected 
to play a significant role in $K\Lambda$ electroproduction \cite{BM00,J02}. 
These resonances do not overlap significantly with the data of Mohring.

While the qualitative discussion above is useful in setting the stage, a
reliable extrapolation of the ${\cal P'}$ polarization data to
$\cos\theta_K^*=1$ is needed to determine $R_\sigma$.  By combining the $z'$
and $z$ components of Eqs.(\ref{eqn_Pxzp}) and (\ref{eqn_Pxz}) and
rearranging, we get:
\begin{equation}
 \label{eqn-Psum1}
  R_{sum}\equiv {{({\cal P}'_{z'}+{\cal P}'_z)\sigma_u}\over c_0}=
     {\cal K}[(1+\cos\theta_K^*)R_{TT'}^{z'0}-R_{TT'}^{x'0}\sin\theta_K^*].
\end{equation}
In this form, both Eqs.(\ref{eqn_Pz1}) and (\ref{eqn-ratio2}) will provide
constraints on $R_{sum}$ at $x\equiv\cos\theta_K^*=\pm1$.  At $x=-1$, the
sum of the polarizations must be zero, according to Eq.(\ref{eqn_Pz1}),
leading to $R_{sum}(x=-1)=0$. This is an important useful constraint that
can most easily be imposed on an extrapolation by using
Eq.(\ref{eqn-Psum1}). Again using the fact that $R_\sigma\ge
0$, Eq.(\ref{eqn-ratio2}) leads to $R_{sum}\leq 2\sigma_u$ for $x=1$.
We will discuss the imposition of these constraints on the extrapolation
shortly. 

Besides the explicit $\theta_K^*$ dependence shown in
Eq.(\ref{eqn-Psum1}) and in the response functions of
Eqs.(\ref{eqn-Rttx}-\ref{eqn-Rl}), the CGLN amplitudes contain additional
$\theta_K^*$ dependence (as well as $Q^2$ and $W$ dependence). This
suggests that Eq.(\ref{eqn-Psum1}) can then be
fit with polynomials in $x=\cos\theta_K^*$ provided we have prior knowledge
of the $\sigma_u$ term.

We have used the unpublished $p(e,e'K^+)\Lambda$ cross section data from
CLAS of Feuerbach~\cite{Feuerbach02} to determine $\sigma_u$. In that work,
the polarization-independent cross sections of Eq.(\ref{csec3}) were
measured over the complete angular range of $\cos\theta^*_K$ and $\Phi$,
and for similar values of $W$ and $Q^2$ as in Ref.~\cite{carman03}.  A
simultaneous fit to the $\cos\theta_K^*$ and $\Phi$ dependence thus enabled
the extraction of the $\cos\theta_K^*$-dependent structure functions of
Eq.(\ref{csec3}): $\sigma_u$, $\sigma_{LT}$, and $\sigma_{TT}$. The cross
section data were fit (8 $\Phi$ bins and 6 $\cos\theta_K^*$ bins per $W$
and $Q^2$ bin) with a third-order polynomial in $x$. The resulting fit
values were then combined with the polarization to calculate the
corresponding values for $R_{sum}$ at the kinematics for each of the data
points in Fig.~\ref{fig-Pz1}.

The number of terms one would include in a polynomial fit to
Eq.(\ref{eqn-Psum1}) is ultimately governed by the reaction dynamics. The
explicit $\theta^*_K$ dependence alone (see
Eqs.(\ref{eqn-Rttx}-\ref{eqn-Rttz})) suggests at least a third-order
polynomial.  However, given the limited number of polarization
data points, the number of terms in any fit leading to a meaningful
extrapolation to $\cos\theta_K^*=1$ must also be limited. We begin by
considering third-order fits of the form:
\begin{equation}
 \label{eqn-Psumfit1}
  R_{sum}= {a_0+a_1x+a_2x^2+a_3x^3}.
\end{equation}

We have done a series of fits to the data points representing $R_{sum}$ in
which we varied the number of terms in the fits while imposing the above
constraints. The fitting routine is a variation of the Levenberg-Marquardt
method~\cite{numrec} of minimizing $\chi^2$. A penalty was imposed on the
$\chi^2$ if a fit strayed too far from the constraints at $x=\pm
1$. Specifically, at $x=-1$, a penalty proportional to the deviation of
$R_{sum}(x=-1)$ from zero was added to the $\chi^2$.  To impose the
constraint at $x=1$, the $\chi^2$ was multiplied by a penalty factor that
was chosen to be large enough to force non-negative values of $R_\sigma$.
In determining the optimal number of parameters in the fit for each $W$
bin, we simply used the number of parameters that produced the smallest
minimized $\chi^2_\nu$ ($\chi^2$ per degree of freedom).

We should point out that the ${\cal P}'_{z'}$ and ${\cal P}'_z$
results of Ref.~\cite{carman03} were obtained from the same data set.
Therefore, these observables are not independent. They do, however, measure
different quantities (as seen by Eqs.(\ref{eqn_Pxzp}) and (\ref{eqn_Pxz}))
since they are projections onto different axes. In adding these together to
form $R_{sum}$, the uncertainties from ${\cal P}'_{z'}$ and ${\cal P}'_z$
were added together.

\begin{figure}[thb]
\vspace{10.6cm}
\includegraphics{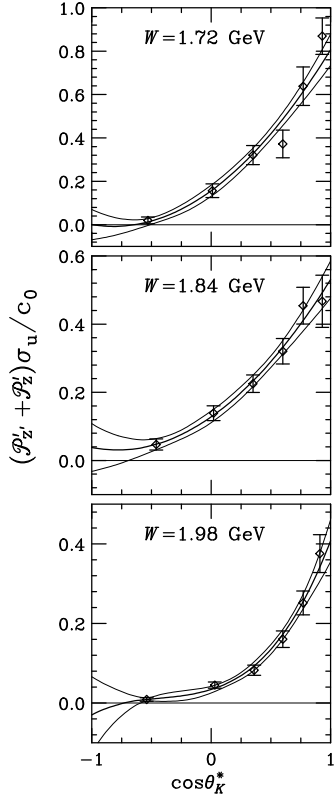} 
\caption[]{$R_{sum}$ (defined in Eq.(\ref{eqn-Psum1})) versus
 $\cos\theta^*_K$ along with our fits (heavy solid lines) and the
 error band resulting from the fit and cross section uncertainties (light 
 solid lines).}
 \label{fig-Rsum}
\end{figure}

For the two lowest $W$ bins we found that a second-order polynomial in $x$
was best for fitting $R_{sum}$ and that a third-order polynomial was
necessary for fitting the highest $W$ bin.  The fact that all $W$ bins do
not require the same order fit should not be surprising since the
underlying physics (CGLN amplitudes) will contribute differently at
different values of $W$.  The results of our fits are shown in
Fig.~\ref{fig-Rsum} (heavy solid lines) along with an error band (light
solid lines). The error bands include uncertainties both from the fitting
of Eq.(\ref{eqn-Psumfit1}), and also contributions from uncertainties in
the fits of the cross section data.  The latter contribution to the
uncertainties is about half of that of the former.  The error band
indicates that the extrapolation to $x=1$ is well constrained but that the
back-angle fit is not.  This is not surprising given the lack of data at
back angles. However, the error bands do encompass the back-angle
constraint that $R_{sum}(x=-1)=0$. Table~\ref{tab-results} shows the
resulting $\chi^2_\nu$ and the polarization extrapolated to $x=1$. The
$\chi^2_\nu$ was determined after removing any penalties remaining in
$\chi^2$. 

We found that the resulting extrapolation of the polarization (and thus
$R_\sigma$) at $x=1$ is relatively insensitive to the exact form of the
unpolarized cross section.  We have done fits in which we varied the
unpolarized cross section used in the fit by 10-20\% (10\% being the stated
upper limit on systematic uncertainty of the unpolarized cross sections
\cite{Feuerbach02}) and observed negligible changes in the resulting value
of $R_\sigma$.  It is the polarization data that dominates the
extrapolation and its associated uncertainties.

\begin{table}[bt]
 \begin{center}
 \begin{tabular}{|c|c|c|c|c|}\hline
 $\langle W\rangle$ GeV & $\langle Q^2\rangle $ GeV$^2$ & $\chi^2_\nu$
 & ${\cal P}'_{z',z}(x=1)$  & $R_\sigma$ \\ \hline\hline
 1.72 & 0.77 & 1.93 &  0.783$\pm$0.072 & 0.005$\pm$0.160$\pm$0.162 \\ \hline
 1.84 & 0.69 & 0.35 &  0.782$\pm$0.091 & 0.239$\pm$0.252$\pm$0.232 \\ \hline
 1.98 & 0.61 & 1.34 &  0.875$\pm$0.080 & 0.088$\pm$0.399$\pm$0.267 \\ \hline
 \end{tabular} 
 \end{center}

 \caption{Transferred polarization at $x=\cos\theta_K^*\!=\! 1.0$
 extrapolated from the fits described in the text along with the resulting
 value of the ratio of transverse to longitudinal structure functions.
 Uncertainties on ${\cal P}'_{z',z}$ are the combined uncertainties arising
 from the fits to both the polarization and cross section data. The first
 uncertainty on $R_\sigma$ is the statistical uncertainty (from the fit)
 while the second represents an estimated systematic uncertainty.}
\label{tab-results}
\end{table}

Plugging the extrapolated polarizations into Eq.(\ref{eqn-ratio2}) we can
determine the ratio $R_\sigma$.  These values are shown in the last column
of Table~\ref{tab-results} along with the combined uncertainties of the
polarization and cross section fits and an estimated systematic
uncertainty. Ref.~\cite{carman03} cites an absolute systematic error of
less than 0.08 for each polarization point. Assuming a comparable
systematic uncertainty for the extrapolated polarization at $x=1$ leads 
to the estimated systematic uncertainties on $R_\sigma$ given in the table.  

\section{Discussion}

\begin{figure}[hbt]
\vspace{6.2cm}
\includegraphics{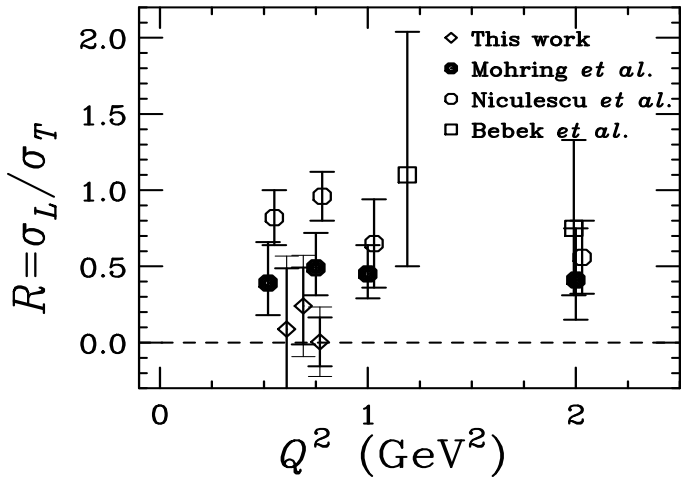} 
 \caption[]{Ratio of longitudinal to transverse structure functions
 vs.~$Q^2$. The inner error bars on our points represent the statistical
 uncertainties arising from the fit and the outer error bars represent the
 combination of statistical and estimated systematic uncertainties. 
 The Niculescu results~\cite{Niculescu}, which were superseded by the
 Mohring results~\cite{Mohring03}, are offset in $Q^2$ for clarity.}
 \label{fig-RLT_ave}
\end{figure}

The resulting values for $R_\sigma$ are plotted in Fig.~\ref{fig-RLT_ave}.
For comparison, we have also included the previously published
data~\cite{Bebek,Niculescu,Mohring03}.  We see that our results for
$R_\sigma$ using polarization transfer data are consistently smaller than
the previous data obtained using the more common Rosenbluth separation
method.  In this method, the unpolarized cross section at a given $Q^2$ and
$W$ is measured in parallel kinematics as a function of $\epsilon$. The
cross section reduces to $\sigma_u=\sigma_T+\epsilon\sigma_L$.  In
principle, fitting the cross section data with a straight line gives
$\sigma_L$ and $\sigma_T$.  However, this technique relies on a very
precise determination of the absolute cross section at all $\epsilon$
points.  To illustrate the inherent difficulties of this technique, one
only needs to compare the results of Niculescu~\cite{Niculescu} with those
of Mohring~\cite{Mohring03}.  These are two analyses of the {\it same} data
set taken in Hall C of Jefferson Lab.  The Mohring results were published
later and are generally believed to be more reliable.  These data were
measured at $W$=1.84~GeV--the same as our middle data point. Our own
results are strikingly different than the Niculescu results and are
slightly lower than--although within uncertainties of--the results of Mohring for
points at comparable kinematics (lower two $Q^2$ points).  We also observe
that our highest $Q^2$ point is consistent with zero to within the
extracted uncertainty.  Again, comparison of this point to the results of
Mohring is probably inappropriate because of the low average value of $W$
for our point.

Our result implies a small longitudinal structure function and hence a
small longitudinal coupling of the virtual photon.  In fact, at the highest
$Q^2$ point (lowest $W$) our result is consistent with $\sigma_L=0$.  This
structure function is expected to be very sensitive to the kaon form
factor~\cite{saghai98}. A recently conducted experiment in Hall A at
Jefferson Laboratory~\cite{markowitz} has as one of its main goals a
Rosenbluth separation at several values of momentum transfer $t$ leading to
a Chew-Low extrapolation~\cite{frazer} of the kaon form factor. However,
this method relies on having small {\it relative} uncertainties for
$\sigma_L$.  Therefore, if $\sigma_L$ is as small as our results indicate,
it seems unlikely that a reliable determination of the kaon form factor can
be performed at these kinematics.

A more significant discrepancy between Rosenbluth and polarization-transfer
results was previously observed in measurements of the ratio of the
proton's electric and magnetic form factors, $\mu_pG_E/G_M$ (see
Ref.~\cite{Arrington1} and references therein).  The common wisdom is that
the polarization-transfer method is much less susceptible to systematic
errors.  If one believes that the polarization transfer results are
correct, then the slope derived from the Rosenbluth technique is too
large. This is similar to what is implied by our results. One explanation
that is being widely discussed is that there is an $\epsilon$-dependent,
two-photon-exchange effect that has not been properly accounted for in the
radiative corrections applied to the experimental cross sections
\cite{Arrington2}. Some calculations~\cite{Blunden} indicate that including
this effect would bring down the high $\epsilon$ data points relative to
the low $\epsilon$ data points and lead to better agreement between the
Rosenbluth and polarization-transfer results.  It is certainly premature to
attribute the small differences between Rosenbluth results and our
polarization-transfer results for $K^+\Lambda$ electroproduction to
two-photon exchange. However, it is interesting that the trend of the
differences is in the same direction as observed in the $\mu_pG_E/G_M$
results.

We would like to note that the polarization transfer data from which we
extracted $R_\sigma$ is just the first to come out of a larger program of
kaon electroproduction in Hall B~\cite{hallb} at Jefferson Lab.  Analysis is
currently underway in which cross sections and polarization observables
will be measured covering $W$ from threshold up to 3.0~GeV and $Q^2$ from
0.3 up to 5~GeV$^2$.  These data will yield a factor of 4 better
statistical uncertainty for the $\Lambda$ polarization transfer, thus
leading to a more reliable determination of $R_\sigma$.  Additional
back-angle data will also allow extraction of $R_\sigma$ in anti-parallel
kinematics.  In addition, Hall B will produce its own Rosenbluth separation
that will complement both the existing data and results that are expected
soon from Hall A~\cite{markowitz}.  We are eager to see if the apparent
small differences between the two techniques of extracting $R_\sigma$ hold 
up with the coming results or are simply a case of statistical 
fluctuations.
 
In conclusion, we have done the first extraction of the ratio
$R_\sigma=\sigma_L/\sigma_T$ from transferred polarization data for the
$p(\vec e,e'K^+)\vec\Lambda$ reaction. Our results are systematically lower
than the results obtained by the Rosenbluth technique and are also
significantly different from model predictions.  These results indicate a
small longitudinal structure function for $Q^2$ of around 0.7~GeV$^2$
and $W$ of 1.72, 1.84, and 1.98~GeV.

\end{document}